\documentclass[a4paper,11pt]{article}
\usepackage{pos}

\def\apj{Astrophys. J.}
\def\apjl{Astrophys. J. Lett.}
\def\apjs{Astrophys. J. Supp.}
\def\mnras{Mon. Not. R. Ast. Soc.}
\def\aap{Astron. Astrophys.}

\title{The luminosity function of TeV-emitting BL Lacs: observations of an HBL sample with VERITAS }
 \ShortTitle{The luminosity function of TeV-emitting BL Lacs}

\author*[a]{M. Errando}

\affiliation[a]{Department of Physics, Washington University in St Louis, St. Louis, MO 63130, USA}

\forColl{VERITAS} 

\emailAdd{errando@physics.wustl.edu}

\abstract{High-frequency-peaked BL Lacs (HBLs) dominate the extragalactic TeV sky, with more than 50 objects detected by the current generation of TeV observatories. Still, the properties of TeV-emitting HBLs as a population are poorly understood due to biases introduced by the observing strategies of Cherenkov Telescopes, limiting our ability to estimate the potential contribution of TeV blazars to the diffuse neutrino, gamma-ray, and cosmic-ray background as well as their role in the late-stage evolution of active galactic nuclei. The VERITAS telescope array has designed a program to quantify and minimize observational biases by selecting a sample of 36 HBLs and measuring their TeV flux at times that are not weighted towards high-flux states. 
Such a survey could form the basis for a measurement of the luminosity function of TeV-emitting HBLs.}

\FullConference{37$^{\rm{th}}$ International Cosmic Ray Conference (ICRC 2021)\\
		July 12th -- 23rd, 2021\\
		Online -- Berlin, Germany}

\begin{document}
\maketitle

\section{The population of TeV-emitting BL~Lacs}
High-frequency-peaked BL Lac-type blazars (HBLs) are the dominant population of extragalactic sources at $E>1$\,TeV. As of June 2021, they constitute at least 53 out of 89 extragalactic sources detected by ground-based gamma-ray telescopes \cite{2008ICRC....3.1341W}. HBLs are likely the main contributors to the total cosmic TeV radiation, with low-frequency-peaked BL Lacs and flat spectrum radio quasars being detected at TeV energies only during short-duration flares, with low duty cycles, and with very soft energy spectra that rarely extend beyond 1 TeV. 

Even though more than 50 HBLs have been detected by current-generation Imaging Atmospheric {Cherenkov} Telescopes (IACTs), the general properties of the population of TeV-emitting HBLs, such as their spatial and luminosity distributions, are largely unconstrained due to the observational biases intrinsic to the operation of IACTs. A leading source of observational bias is the lack of sensitive blind surveys of the extragalactic TeV sky. These cannot be efficiently conducted with the current generation of IACTs due to their narrow field of view and limited sensitivity. Blazars with TeV fluxes of the order of 0.01\,Crab still take $\gtrsim 25$\,h of exposure to detect.  In addition, observations of TeV blazars are often triggered by flaring states, with the reported TeV fluxes not representing the average TeV emission from a source. Water Cherenkov observatories are not affected by the same sources of observational biases as IACTs thanks to their wide instantaneous field of view and large duty cycle, but with the current sensitivity level of the HAWC survey only two nearby blazars have been detected \cite{2020ApJ...905...76A}.

A measurement of the luminosity function of any class of sources, i.e. the number of sources per unit volume per unit luminosity, is key to study their properties, their relationship with other source classes, and their contribution  to unresolved radiation fields. In addition, the luminosity function of TeV-emitting HBLs
can be used to constrain the intensity of the intergalactic magnetic fields (IGMF), to understand if BL Lac-type blazars are plausible sources of astrophysical neutrinos, and to test models for AGN evolution.

Production of multi-TeV neutrinos in astrophysical sources must be accompanied by production of gamma rays. Barring opacity generated at the source or in intergalactic space, gamma-rays produced in these hadronic processes would point to neutrino sources. The ratio of gamma/neutrino luminosities depends only on the ratio of neutral/charged pions in $pp$ and $p-\gamma$ interactions \cite{2018PrPNP.102...73A}. Most current efforts to link astrophysical neutrinos with BL Lacs are based on a handful of IceCube neutrino events that are spatially, and some times temporally coincident with gamma-ray activity of a BL Lac-type blazar \cite{2018Sci...361.1378I,2018ApJ...861L..20A}. While these efforts must be continued, they do not make use of most of the 650,000 candidate neutrino events \cite{2019ICRC...36.1017S} collected by IceCube that characterize the diffuse neutrino flux and its energy spectrum.  Knowing the global properties of the TeV emission from HBLs (by measuring their luminosity function) will place a constraint on 
their contribution to the extragalactic GeV background, which is the analog of the diffuse neutrino fluxes currently measured by IceCube. Extrapolating the energy density of this contribution to that of the diffuse high-energy neutrino flux using a power-law emission model provides an independent, population-based way to estimate the extent to which HBL blazar jets produce high-energy neutrinos \cite{2017ApJ...835...45A,2019ApJ...871...41P,2020JETP..131..265N}. 
By using the full IceCube neutrino data set and a complete set of TeV observations, such studies will be more robust and constraining than 
current studies limited to small subsamples of TeV and neutrino data.

TeV emission from blazar jets is reprocessed into GeV radiation through pair production on the extragalactic background light (EBL) and the cosmic microwave background (CMB)  \cite{1994ApJ...423L...5A}. In the presence of a non-zero IGMF, this process can lead to the appearance of extended GeV halos around AGN, although these have so far not been detected \cite{2018ApJS..237...32A}. If the intensity of the IGMF is strong, GeV emission produced by pair cascades would be completely isotropized and appear as a component of the isotropic diffuse GeV emission measured by {\it Fermi}-LAT \cite{2015ApJ...799...86A}. Quantitative knowledge of the number density and intensity of TeV jets, through their luminosity function, combined with our current knowledge on the EBL density, allows a simple estimation of how much diffuse GeV emission is expected to arise from pair cascades, with the intensity of the IGMF being the only variable. This has the potential to dramatically improve current constraints on the intensity of the IGMF with a measurement that has different systematic uncertainties than searches for extended halos, increasing our understanding of the cosmological impact of TeV blazar emission. 

Flat spectrum radio quasars (FSRQs) display a positive redshift evolution at all measured energy bands, indicating that there were more FSRQs in the past, up to a cutoff redshift \cite{1990MNRAS.247...19D,2012ApJ...751..108A}. The situation for BL~Lacs is less clear, with some samples showing positive, flat, and negative redshift evolution \cite{2003AA...401..927B,2014ApJ...780...73A}. A possible interpretation is that BL Lacs represent an accretion-starved end-state of an earlier merger-driven gas-rich phase (FSRQs). 
Under this model, BL Lacs would evolve further into TeV BL Lacs (HBLs) where particles can be accelerated to the highest energies due to reduced radiative cooling. This scenario can be confirmed even on small source samples using the $\left< V/V_{\rm max}\right>$ technique \cite{1968ApJ...151..393S,1980ApJ...235..694A} if TeV-selected HBLs show a more negative evolution than LAT-selected BL Lacs.
In addition, the ratio of the TeV and X-ray luminosity functions of HBLs can be interpreted as a measurement of the prevalence of TeV emission in relativistic jets. This would allow us to answer the question of how common the presence of TeV emission is, 
and shed light on the conditions under which particle acceleration to multi-TeV energies can happen.


\section{The VERITAS HBL sample}
The most robust way to measure the luminosity function of TeV-emitting blazars would be to conduct a blind survey of the extragalactic sky \cite{2013APh....43..215S,2013APh....43..317D}. However, the population of blazars detected by the current generation of IACTs has a spatial density of $\sim 2\times 10^{-3}\,\textrm{deg}^{-2}$. This indicates that unless a large population of bright blazars has eluded detection, the odds of finding a new detectable blazar with a 25\,h VERITAS exposure towards a random extragalactic direction are of the order of $\lesssim 1/100$. 

\begin{figure}[t]
\centering
 \includegraphics[width=0.49\linewidth]{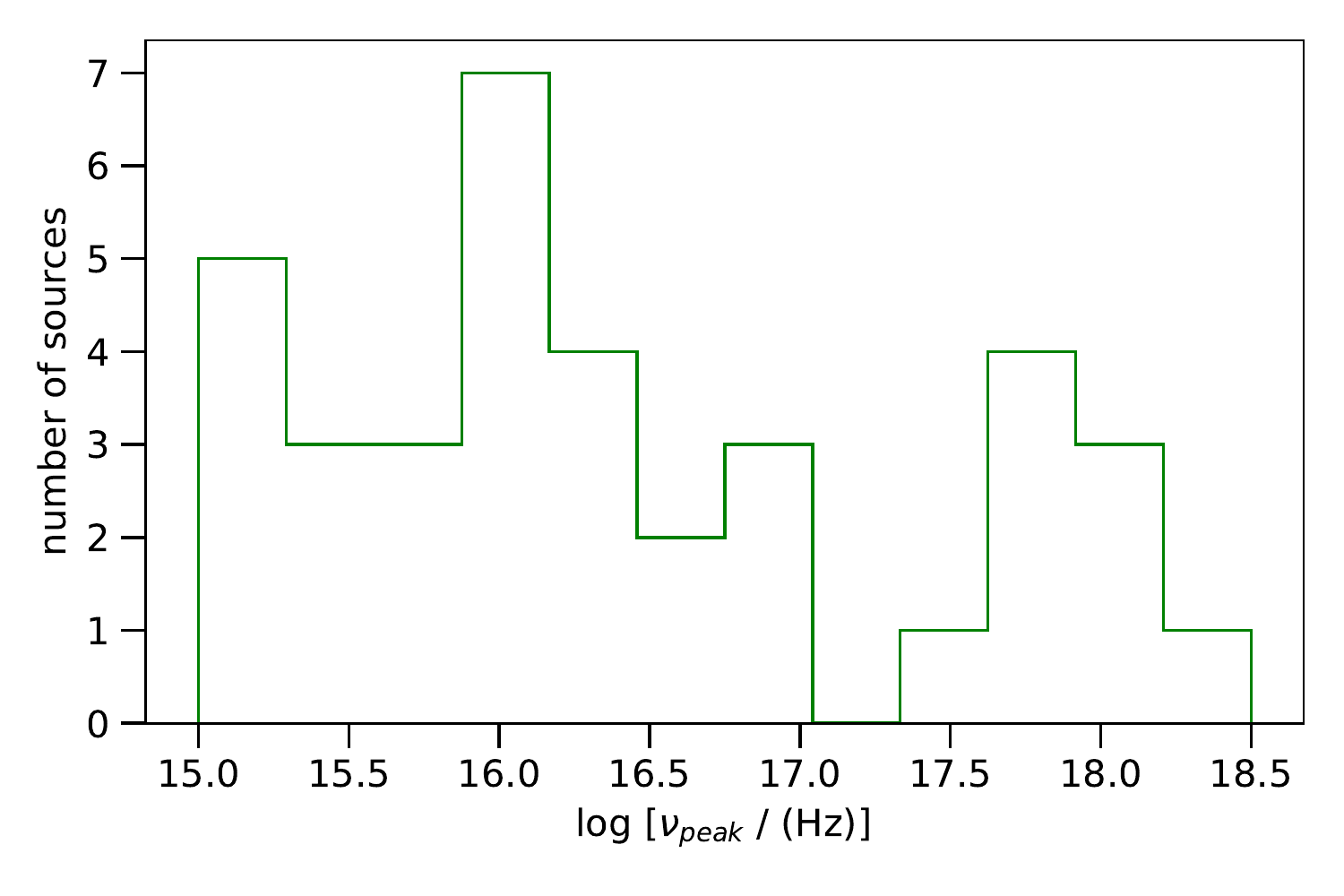}    
 \hspace{0.cm}
\includegraphics[width=0.49\linewidth]{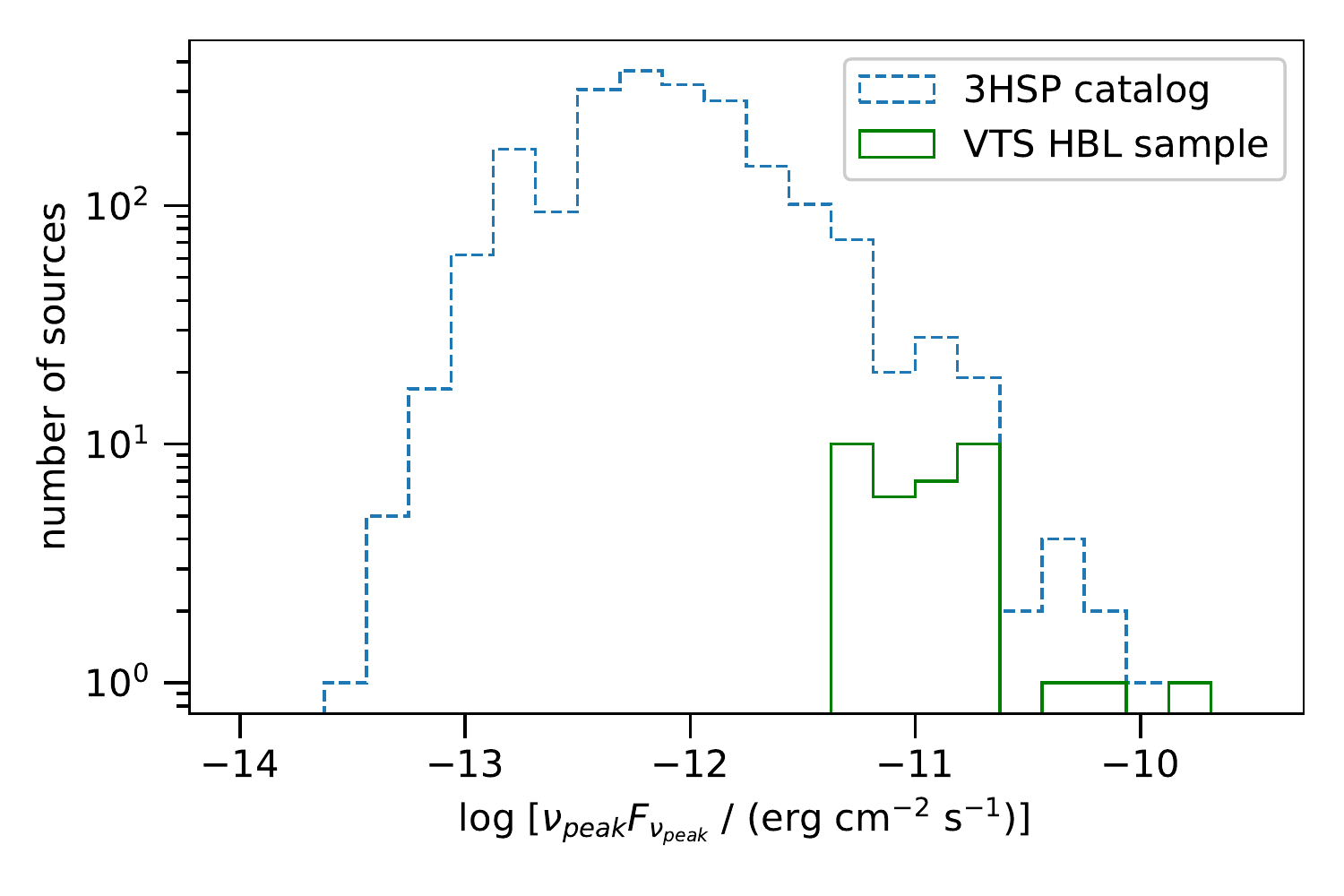}    
\caption{Distribution of the frequency where the synchrotron peak is located ($\nu_{\textrm{peak}}$) and the flux at the synchrotron peak ($\nu_\textrm{peak}F_{\nu_\textrm{peak}}$) as estimated in the 3HSP catalog \cite{2019AA...632A..77C} for the VERITAS HBL sample. }
    \label{fig:nupeak}
\end{figure}

\begin{figure}[t]
\centering
 \includegraphics[width=0.75\linewidth]{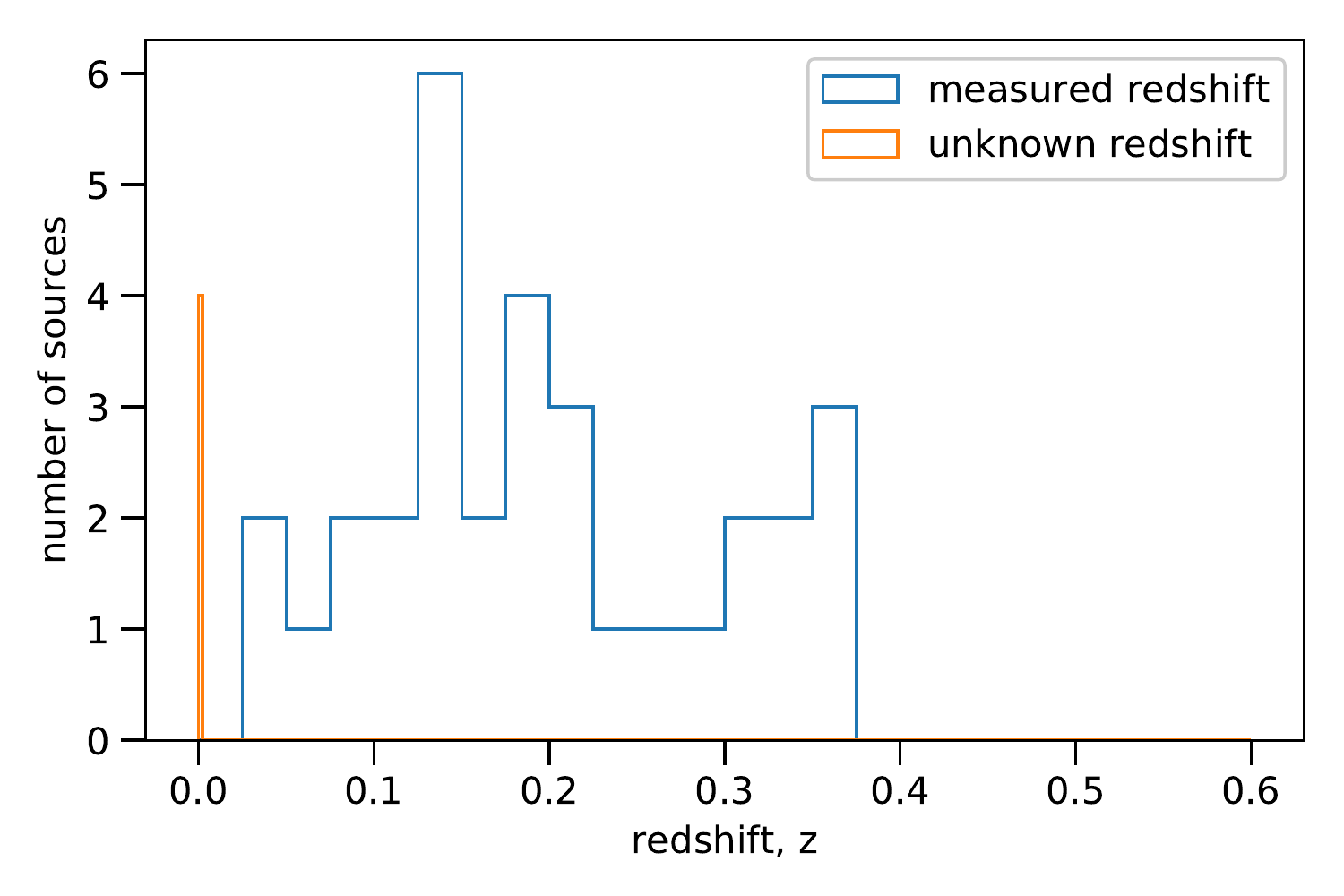}    
\caption{Redshift distribution of the sources in the VERITAS HBL sample. Sources with uncertain redshift can be incorporated into the calculation of the luminosity function by assuming a probability density function for their redshift that takes into account redshift constraints \cite{2014ApJ...780...73A}.}
    \label{fig:z}
\end{figure}

A more efficient way to set constraints on the  luminosity function of HBLs is to measure the TeV flux of a complete sample of TeV-emitting HBLs. We started our selection with the 2WHSP catalog \cite{2017AA...598A..17C}, which cross-correlates sources with infrared spectra similar to known TeV blazars with radio and X-ray data to calculate the location and flux density of the synchrotron peak of the blazars. The 2WHSP catalog was recently superseeded by the 3HSP catalog \cite{2019AA...632A..77C}, which uses the same technique but updates multiwavelength data sets. The VERITAS HBL sample comprises 36 blazars selected according to the following criteria:
\begin{itemize}
\item Object is listed in the 3HSP catalog, indicating $\log(\nu_\textrm{peak}\,\textrm{Hz} > 15)$, i.e. a synchrotron peak in the ultraviolet to X-ray range (see Figure~\ref{fig:nupeak}). 
\item Object is at declination $1.7^\circ \leq {\rm decl.} \leq 61.7^\circ$ to guarantee good observing conditions with VERITAS. 
\item Object is at galactic latitude $\left| {b} \right|>10^\circ$ to avoid incompleteness in seed catalogs near the galactic plane. 
\item Object has an estimated flux at the synchrotron peak of $\log[ \nu_\textrm{peak}F _{\nu_\textrm{peak}} / (\textrm{erg cm}^{-2} \textrm{s}^{-1})] > -11.2$ to focus observing efforts on sources with potential to be detected with VERITAS in $\lesssim 25$\,h of exposure (see Figure~\ref{fig:nupeak}). 
\end{itemize}
The above selection criteria result in 36 northern HBL objects listed in Table~\ref{vhs}, 21 of which are reported in TeVCat \cite{2008ICRC....3.1341W}. The sources span a range of redshift $0.03 < z \lesssim 0.36$, with four sources having unknown or uncertain redshift (Figure~\ref{fig:z}).

\begin{table}[b]
\center
\begin{tabular}{lcccc}
\hline\hline
Object & RA (J2000) & decl. (J2000) & z & TeVCat?\\
\hline
1ES 0120+340 & 01:23:08.6 & +34:20:48.5 & 0.270 &   \\
RGB J0136+391 & 01:36:32.6 & +39:05:59.2 &  & Y \\
RGB J0152+017 & 01:52:39.6 & +01:47:17.4 & 0.080 & Y \\
1ES 0229+200 & 02:32:48.6 & +20:17:17.3 & 0.139 & Y \\
RGB J0316+090 & 03:16:12.7 & +09:04:43.2 & 0.372 &   \\
1FGL J0333.7+2919 & 03:33:49.0 & +29:16:31.5 &   &   \\
GB6 J0540+5823 & 05:40:30.0 & +58:23:38.4 &  &   \\
1ES 0647+250 & 06:50:46.5 & +25:02:59.5 & 0.203 & Y \\
RGB J0710+591 & 07:10:30.1 & +59:08:20.5 & 0.120 & Y \\
PGC 2402248 & 07:33:26.8 & +51:53:55.9 & 0.090 & Y \\
1ES 0806+524 & 08:09:49.2 & +52:18:58.3 & 0.137 & Y \\
87GB 083437.4+150850 & 08:37:24.6 & +14:58:20.6 & 0.278 &   \\
RGB J0847+115 & 08:47:12.9 & +11:33:50.2 & 0.198 & Y \\
RX J0910.6+3329 & 09:10:37.0 & +33:29:24.4 & 0.350 &   \\
B2 0912+29 & 09:15:52.4 & +29:33:24.0 & 0.190 &   \\
1ES 1011+496 & 10:15:04.1 & +49:26:00.8 & 0.200 & Y \\
1ES 1028+511 & 10:31:18.5 & +50:53:35.9 & 0.360 &   \\
RGB J1037+571 & 10:37:44.3 & +57:11:55.7 & 0.330&   \\
RGB J1058+564 & 10:58:37.7 & +56:28:11.2 & 0.143 &   \\
Mrk 421 & 11:04:27.3 & +38:12:31.9 & 0.030& Y \\
RX 1117.1+2014 & 11:17:06.3 & +20:14:07.5 & 0.138 &   \\
1ES 1218+304 & 12:21:22.0 & +30:10:37.2 & 0.180 & Y \\
MS 1221.8+2452 & 12:24:24.2 & +24:36:23.6 & 0.218 & Y \\
S3 1227+25 & 12:30:14.1 & +25:18:07.1 & 0.135 & Y \\
RGB J1243+364 & 12:43:12.7 & +36:27:44.0 & 0.310 &   \\
RBS 1366 & 14:17:56.7 & +25:43:25.9 & 0.240 &   \\
H 1426+428 & 14:28:32.6 & +42:40:21.0 & 0.129 & Y \\
RGB J1439+395 & 14:39:17.5 & +39:32:42.8 & 0.344 &   \\
1ES 1440+122 & 14:42:48.2 & +12:00:40.3 & 0.160 & Y \\
PG 1553+113 & 15:55:43.0 & +11:11:24.4 & 0.360 & Y \\
Mrk 501 & 16:53:52.2 & +39:45:36.5 & 0.030 & Y \\
H 1722+119 & 17:25:04.3 & +11:52:15.5 & 0.180 & Y \\
1ES 1727+502 & 17:28:18.6 & +50:13:10.5 & 0.055 & Y \\
RGB J1838+480& 18:38:49.1 & +48:02:34.4 & 0.300 &   \\
RGB J2243+203 & 22:43:54.7 & +20:21:03.8 &   & Y \\
B3 2247+381 & 22:50:05.7 & +38:24:37.2 & 0.119 & Y \\
\hline
\hline
\end{tabular}
\caption{The VERITAS HBL sample. Some of the quoted redshifts are uncertain. }
\label{vhs}
\end{table}

\section{VERITAS observations}
The VERITAS observatory \citep{VERITAST1,VERITAS} is an array of four
imaging atmospheric Cherenkov telescopes located at the Fred Lawrence Whipple Observatory in southern Arizona (31$^\circ$ 40$^\prime$ N, 110$^\circ$ 57$^\prime$ W,  1.3\,km a.s.l.). Each telescope consists of a 12\,m diameter reflector and a photomultiplier camera 
covering a field of view of $3.5^{\circ}$.
The array has an effective area of $\sim 5 \times 10^4\,\mathrm{m}^{2}$ between 0.2 and 
10\,TeV.  

VERITAS detects a source with 1\% of the Crab Nebula 
flux\footnote{$1\,\mathrm{Crab}=2.1\times
10^{-10}\mathrm{cm}^{-2}\mathrm{s}^{-1}$ at $E>0.2$\,TeV \citep[][]{hillas-crab}} in $\sim 25$\,h of 
exposure, covering the energy range between 0.1\,TeV to $>30$\,TeV.
The angular and energy resolution for reconstructed gamma-ray showers are $\sim 0.1 ^{\circ}$ and 
15\%, respectively, at 1\,TeV.

Since the start of  four-telescope operations in 2007, VERITAS has collected more than 2000 hours of exposure on the 36 sources that form the VERITAS HBL sample. This includes 155\,h of dedicated observations between 2019 and 2021 that were obtained in order to achieve a sensitivity of $\sim 1\%$ of the Crab Nebula flux for all objects. To avoid biasing the flux measurements towards flaring states, archival data were filtered by looking at observation logs and removing VERITAS observations that were triggered by high flux states detected at other wavelengths (optical, X-ray, GeV), alerts and ATel notifications from other TeV observatories, or self-triggered by high flux states or signal excesses measured with VERITAS or the Whipple 10-m telescope. This typically results in the exclusion of $\sim 30\%$ of the data in the VERITAS archive, although that figure varies significantly from source to source. 


Analysis of the VERITAS data obtained so far is underway, with all additional observations expected to be completed in the 2021-22 season \cite{2019ICRC...36..638B}. Once flux measurements or upper limits have been derived for all 36 sources in the VERITAS HBL sample, constraints on the luminosity function can be derived by estimating, via simulations, the completeness of the VERITAS survey and that of the seed 3HSP catalog, following established techniques \cite{2014ApJ...780...73A}. Simulation work indicates that VERITAS will be able to constrain the spectral index of a power-law luminosity function with an expected uncertainty of 30\%, and would be will be sensitive to potential redshift evolution by detecting 10\%-level deviations from 0.5 in $\langle V/V_\textrm{max}\rangle$ \cite{ari}.

Results from the distribution of TeV fluxes from the VERITAS HBL sample and the resulting measurement of the luminosity function will provide constraints to the total amount of TeV radiation produced by HBLs, informing future studies with the Cherenkov Telescope Array. While HBLs are the dominant class of extragalactic TeV sources, VERITAS is also conducting studies to constrain the contribution from low- and intermediate-frequency peaked BL~Lacs and flat spectrum radio quasars \cite{sonalICRC} to the cosmic TeV flux from blazars.

\section*{Acknowledgments}
This research is supported by grants from the U.S. Department of Energy Office of Science, the U.S. National Science Foundation and the Smithsonian Institution, by NSERC in Canada, and by the Helmholtz Association in Germany. This research used resources provided by the Open Science Grid, which is supported by the National Science Foundation and the U.S. Department of Energy's Office of Science, and resources of the National Energy Research Scientific Computing Center (NERSC), a U.S. Department of Energy Office of Science User Facility operated under Contract No. DE-AC02-05CH11231. We acknowledge the excellent work of the technical support staff at the Fred Lawrence Whipple Observatory and at the collaborating institutions in the construction and operation of the instrument.

\clearpage
\section*{Full Authors List: \Coll\ Collaboration}

\scriptsize
\noindent
C.~B.~Adams$^{1}$,
A.~Archer$^{2}$,
W.~Benbow$^{3}$,
A.~Brill$^{1}$,
J.~H.~Buckley$^{4}$,
M.~Capasso$^{5}$,
J.~L.~Christiansen$^{6}$,
A.~J.~Chromey$^{7}$, 
M.~Errando$^{4}$,
A.~Falcone$^{8}$,
K.~A.~Farrell$^{9}$,
Q.~Feng$^{5}$,
G.~M.~Foote$^{10}$,
L.~Fortson$^{11}$,
A.~Furniss$^{12}$,
A.~Gent$^{13}$,
G.~H.~Gillanders$^{14}$,
C.~Giuri$^{15}$,
O.~Gueta$^{15}$,
D.~Hanna$^{16}$,
O.~Hervet$^{17}$,
J.~Holder$^{10}$,
B.~Hona$^{18}$,
T.~B.~Humensky$^{1}$,
W.~Jin$^{19}$,
P.~Kaaret$^{20}$,
M.~Kertzman$^{2}$,
T.~K.~Kleiner$^{15}$,
S.~Kumar$^{16}$,
M.~J.~Lang$^{14}$,
M.~Lundy$^{16}$,
G.~Maier$^{15}$,
C.~E~McGrath$^{9}$,
P.~Moriarty$^{14}$,
R.~Mukherjee$^{5}$,
D.~Nieto$^{21}$,
M.~Nievas-Rosillo$^{15}$,
S.~O'Brien$^{16}$,
R.~A.~Ong$^{22}$,
A.~N.~Otte$^{13}$,
S.~R. Patel$^{15}$,
S.~Patel$^{20}$,
K.~Pfrang$^{15}$,
M.~Pohl$^{23,15}$,
R.~R.~Prado$^{15}$,
E.~Pueschel$^{15}$,
J.~Quinn$^{9}$,
K.~Ragan$^{16}$,
P.~T.~Reynolds$^{24}$,
D.~Ribeiro$^{1}$,
E.~Roache$^{3}$,
J.~L.~Ryan$^{22}$,
I.~Sadeh$^{15}$,
M.~Santander$^{19}$,
G.~H.~Sembroski$^{25}$,
R.~Shang$^{22}$,
D.~Tak$^{15}$,
V.~V.~Vassiliev$^{22}$,
A.~Weinstein$^{7}$,
D.~A.~Williams$^{17}$,
and 
T.~J.~Williamson$^{10}$\\
\noindent
$^{1}${Physics Department, Columbia University, New York, NY 10027, USA}
$^{2}${Department of Physics and Astronomy, DePauw University, Greencastle, IN 46135-0037, USA}
$^{3}${Center for Astrophysics $|$ Harvard \& Smithsonian, Cambridge, MA 02138, USA}
$^{4}${Department of Physics, Washington University, St. Louis, MO 63130, USA}
$^{5}${Department of Physics and Astronomy, Barnard College, Columbia University, NY 10027, USA}
$^{6}${Physics Department, California Polytechnic State University, San Luis Obispo, CA 94307, USA} 
$^{7}${Department of Physics and Astronomy, Iowa State University, Ames, IA 50011, USA}
$^{8}${Department of Astronomy and Astrophysics, 525 Davey Lab, Pennsylvania State University, University Park, PA 16802, USA}
$^{9}${School of Physics, University College Dublin, Belfield, Dublin 4, Ireland}
$^{10}${Department of Physics and Astronomy and the Bartol Research Institute, University of Delaware, Newark, DE 19716, USA}
$^{11}${School of Physics and Astronomy, University of Minnesota, Minneapolis, MN 55455, USA}
$^{12}${Department of Physics, California State University - East Bay, Hayward, CA 94542, USA}
$^{13}${School of Physics and Center for Relativistic Astrophysics, Georgia Institute of Technology, 837 State Street NW, Atlanta, GA 30332-0430}
$^{14}${School of Physics, National University of Ireland Galway, University Road, Galway, Ireland}
$^{15}${DESY, Platanenallee 6, 15738 Zeuthen, Germany}
$^{16}${Physics Department, McGill University, Montreal, QC H3A 2T8, Canada}
$^{17}${Santa Cruz Institute for Particle Physics and Department of Physics, University of California, Santa Cruz, CA 95064, USA}
$^{18}${Department of Physics and Astronomy, University of Utah, Salt Lake City, UT 84112, USA}
$^{19}${Department of Physics and Astronomy, University of Alabama, Tuscaloosa, AL 35487, USA}
$^{20}${Department of Physics and Astronomy, University of Iowa, Van Allen Hall, Iowa City, IA 52242, USA}
$^{21}${Institute of Particle and Cosmos Physics, Universidad Complutense de Madrid, 28040 Madrid, Spain}
$^{22}${Department of Physics and Astronomy, University of California, Los Angeles, CA 90095, USA}
$^{23}${Institute of Physics and Astronomy, University of Potsdam, 14476 Potsdam-Golm, Germany}
$^{24}${Department of Physical Sciences, Munster Technological University, Bishopstown, Cork, T12 P928, Ireland}
$^{25}${Department of Physics and Astronomy, Purdue University, West Lafayette, IN 47907, USA}


%
%
%

\end{document}